\documentclass[12pt,preprint]{aastex}

\shorttitle{Characteristic Dependence of Umbral Dots}
\shortauthors{H.Watanabe et al.}

\begin{document}

\title{Characteristic Dependence of Umbral Dots on their Magnetic Structure}

\author{H. Watanabe, R. Kitai, and K. Ichimoto}
\affil{Kwasan and Hida Observatories, Kyoto University, Yamashina-ku, Kyoto 607-8417, JAPAN}
\email{watanabe@kwasan.kyoto-u.ac.jp}

\altaffiltext{1}{Kwasan and Hida Observatories, Kyoto University, Yamashina-ku, Kyoto 607-8417, JAPAN}

\begin{abstract}
Umbral dots (UDs) were observed in a stable sunspot in NOAA 10944 by the \emph{Hinode} Solar Optical Telescope on 2007 March 1.
The observation program consisted of blue continuum images and spectropolarimetric profiles of Fe {\footnotesize I} 630\,nm line. 
An automatic detection algorithm for UDs was applied to the 2-hour continuous blue continuum images, and using the obtained data, the lifetime, size, and proper motion of UDs were calculated. 
The magnetic structure of the sunspot was derived through the inversion of the spectropolarimetric profiles.  

We calculated the correlations between UD's parameters (size, lifetime, occurrence rate, proper motion) and magnetic fields (field strength, inclination, azimuth), and obtained the following results: (1) Both the lifetime and size of UDs are almost constant regardless of the magnetic field strength at their emergence site.  (2) The speed of UDs increases as the field inclination angle at their emergence site gets larger. (3) The direction of movement of UDs is nearly parallel to the direction of the horizontal component of magnetic field in the region with strongly inclined field, while UDs in the region with weakly inclined field show virtually no proper motion.

Our results describe the basic properties of magnetoconvection in sunspots.  We will discuss our results in comparison to recent MHD simulations by \cite{Schussler2006} and \cite{Rempel2009}.  
\end{abstract}


\section{Introduction}
Umbral dots (UDs), small bright points with size of $\sim$300\,km and lifetime of $\sim$15\,min, are observed ubiquitously across the umbra \citep{Sobotka1997a, Sobotka1997b, Thomas2004}.
Since the first observation of UDs by \citet{Chevalier1916}, it has been found that UDs play an important role in the energy balance in sunspots.  
The brightness of an umbra is about 5-20\% of that of a quiet region, because convection is strongly suppressed in the presence of a strong magnetic field (a few thousand Gauss in an umbra).
However, in order to account for the observed brightness of an umbra, heat must be supplied by convective transport \citep{Deinzer1965}.
UDs are the signature of this convective heat transport. 
Therefore, understanding the physics of UDs is key to constructing a precise model of sunspot substructure.

The mechanism behind the formation of UDs is thought to be magnetoconvection.
In recent years, theoretical studies of magnetoconvection have greatly improved knowledge of the phenomena.
Two-dimensional magnetoconvection in a Boussinesq fluid was studied by \cite{Proctor1982}, and a systematic investigation of compressible magnetoconvection was described by \cite{Weiss1990}.  
\cite{Weiss1990} revealed that the modes of magnetoconvection are governed by the ratio $\zeta$ of the magnetic diffusivity to the thermal diffusivity.
For $\zeta \lesssim 1$, we obtain oscillatory convection with periodic reversals of the flow velocity.
For $\zeta \gtrsim 1$, overturning convection occurs with a spatially asymmetric rising and falling of the convective plumes.
In the quiet photosphere, the thermal diffusivity usually exceeds the magnetic diffusivity and as a result $\zeta$ is far smaller than unity. 
In the umbra, however, the atmospheric conditions are such that $\zeta < 1$ near the surface and $\zeta > 1$ at depth below 1500\,km.
Because of these complex atmospheric condition, the magnetoconvection in sunspots has been investigated mainly through the use of computer simulations \citep{Schussler2006, Heinemann2007, Rempel2009}.
In the simulations by \citet{Schussler2006}, they found that UDs are convective plumes that are triggered by oscillatory convection, but turn into overturning cells because of the radiative cooling at the surface.  
Upward velocity surrounded by downflow was found in Fe {\footnotesize I} 5576{\AA} line observation \citep{Bharti2007}, which was suggested to be supporting evidence of overturning convection.

Observations of umbra have revealed that UDs form at regions with a reduced field strength and a local upward velocity \citep{Watanabe2009, Sobotka2009}. 
\citet{SocasNavarro2004} found that there are systematic differences between UDs and the surrounding umbra, such as small upflows ($\sim$100 m s$^{-1}$), higher temperatures ($\sim$1 kK), and weaker fields ($\sim$500\,Gauss) with more inclined orientations ($\sim$10$^{\circ}$).
These results are consistent with theoretical models for the convection processes driving UDs.
In this paper, we report the statistical relationship between the magnetic field and UDs through the use of filtergram and spectrogram observations from the \emph{Hinode} Solar Optical Telescope (SOT). 
If UDs are the convective plumes controlled by the magnetic field, there must be correlations between UDs and magnetic field, for example, between field strength and UD's size, or between field strength and proper motion of UDs.
Our results can serve as a guide for more realistic MHD simulations, leading toward an ultimate understanding of the subsurface structure of sunspots.

In the following sections, we describe the observations and the data reduction techniques (\S2), explain the UD detection algorithm (\S3), analyze the properties of UDs and their relationship to the magnetic fields (\S4), and finally we will discuss the results in \S5.  

\section{Observation and Data Reduction}
The target in this paper is a stable and circular sunspot in NOAA 10944 observed with the {\it Hinode} SOT.
With the SOT/Broadband Filter Imager (BFI) \citep{Tsuneta2008}, blue continuum images were taken from 00:14UT till 02:30UT on Mar 1, 2007.
The blue continuum image by the SOT/BFI has wavelength centered on 450.45\,nm and the line width of 0.4\,nm. 
The diffraction limit of blue continuum for the {\it Hinode} SOT is 164\,km.
On Mar 1, the sunspot was located almost at the disk center:
the heliocentric coordinates for the sunspot at 00:14UT on Mar 1 were (63{\arcsec}, 17{\arcsec}).
The images were taken with a constant 6\,s interval.
However, since we analyzed every 4th image, in our study the temporal cadence is $\sim$25\,s.
The spatial pixel size was 0.054{\arcsec} $\times$ 0.054{\arcsec}, and the exposure time was relatively long, 102\,ms, in order to obtain adequate photons for the umbra.
The field-of-view (FOV) of the images was 54{\arcsec} $\times$ 27{\arcsec}, which contains the entire umbra.

The total number of the continuum images was 321, which covers the observational period of 136\,min.
After the dark field subtraction and the flat fielding, the images were carefully co-aligned by finding the displacement which gave the maximum cross-correlation between consecutive frames.
In order to reduce high-frequency noise in space caused by CCD or photon noise, we applied a lowpass filter (hanning filter) to the Fourier transformed image, and then an inverse Fourier transformation was performed.
Finally, we normalized the images with a low-frequency component of the averaged lightcurve of the quiet-sun region.
In doing so, we removed the effect of the small orbital variation of the CCD gain whose period is $\sim$96 min.  

In addition to the filtergram, we used the \emph{Hinode} spectropolarimeter (SP) data to study the magnetic field of the sunspot. 
Unfortunately, no SP data was taken simultaneously with the blue continuum imaging.
Instead, we created a composite map from the two closest SP maps taken before and after the filtergram imaging. 
One map was taken at 17:58UT on Feb 28 (about 6 hours before the start of the filtergram observation), and the other was taken at 06:14 on Mar 1 (about 4 hours after the end of the filtergram observation).
The spectral FOV includes the two magnetic-sensitive iron lines, Fe {\footnotesize I} 630.15\,nm and  630.25\,nm.
The two SP scans were carried out in normal mode.
In this mode, it takes about 45 min to scan the FOV of 76{\arcsec} $\times$ 82{\arcsec} with a polarization accuracy of 0.1\%.
The observed full Stokes parameters ({\it I,Q,U,V}) were processed through a dark field subtraction, a flat fielding, and a thermal drift calibration using the standard routine. 

To extract the magnetic field information, we applied the Milne-Eddington inversion (Yokoyama et al. 2009, in preparation) to the calibrated profiles.
The umbral photosphere is well suited to the application of Milne-Eddington inversions, since the velocity and magnetic field gradients are very small.
As a result of the inversion, we obtained maps of magnetic field strength ($B$), field inclination ({$i$), and field azimuth ($\psi$).

The two magnetic maps, taken at the different times, were rotated to match the sunspot's orientation at 01:30UT on Mar 1 (in the middle of the filtergram observation) by a coordinate rotation.
In this procedure, the magnetic field inclination and azimuth were converted using a planar approximation.
The inclination is, hereafter, defined by the angle between a field line and the local normal, i.e., 0$^{\circ}$ inclination means a vertical, 90$^{\circ}$ inclination means a horizontal field line.
The field azimuth is measured counter-clockwise from the right-to-left (east) direction.
Subsequently, the two maps were co-aligned by finding the best cross-correlation displacements and averaged into one map.
Finally, the magnetic strength and inclination maps were boxcar-smoothed with a width of 1.1{\arcsec} $\times$ 1.1{\arcsec} to extract the global structure.
We define this composite map as the magnetic field at 01:30UT on Mar 1, and use it for the following analysis.
It is worth noting that our composite magnetic map can be used only for retrieving the global magnetic characteristics, because the individual original magnetic maps (6 hours before and 4 hours afterwards) may include local variations contributed by UDs.

The obtained SP map is enlarged to have the same pixel size as that of the blue continuum image.  
For the alignment process between the blue continuum and the SP, we made the line wing intensity map at around 630.3\,nm. 
The smallest displacement was calculated using the blue continuum image taken at 01:30UT and the line wing intensity map by a cross-correlation analysis.

As can be seen in Fig.\,\ref{fig:SPbefore_after}, the sunspot was almost circular and encompassed with a penumbra.
Two umbral dark core regions with large field strengths $>$2600\,Gauss can be seen in the southern region of the sunspot.
Between these two dark cores, a short light bridge connects to the southern penumbra .

\begin{figure}[bhtp]
\epsscale{1.0}
\plotone{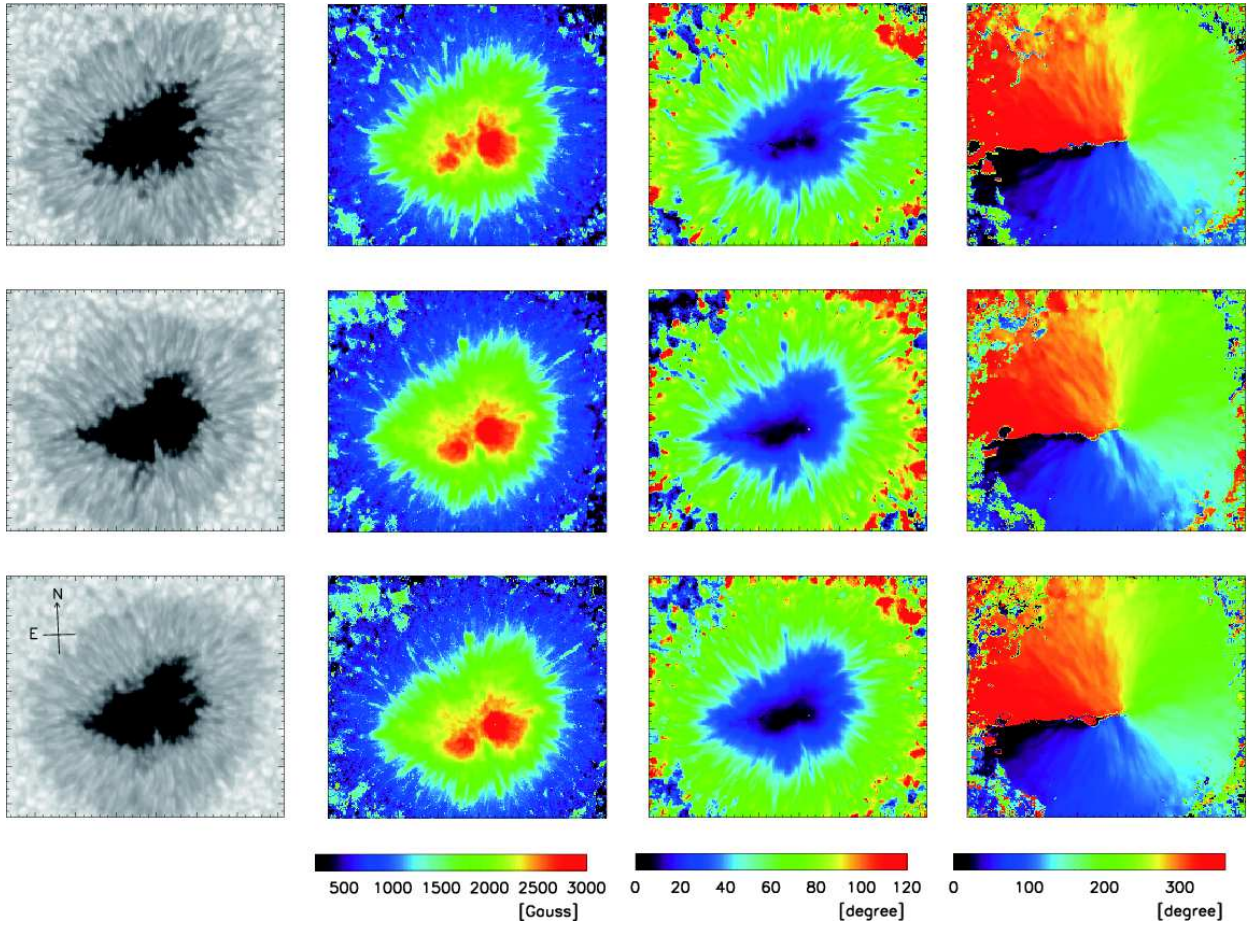}
   \caption{From left to right: the continuum intensity at $\sim$630.3\,nm, magnetic field strength, field inclination, and field azimuth. From top column to bottom: the SP at 17:58UT on Feb 28 (the images were rotated to the sunspot's location at 01:30UT on Mar 1, see text), the SP at 06:14UT on Mar 1 (ditto), and the composite map of the two (before smoothing). The marks are shown in 1{\arcsec} intervals on all sides of the images.}
    \label{fig:SPbefore_after}
\end{figure}

\section{Automatic Detection Algorithm}
To perform a statistical analysis of UDs, it is necessary to use an automatic detection algorithm, because the sunspot of interest produced numerous UDs.
Based on previous papers which utilized automated algorithms, such as \cite{Sobotka1997a} and \cite{Riethmuller2008}, we constructed a new automatic detection algorithm.
Compared to previous work, our algorithm can be kept simpler since the {\it Hinode} data is free from variable atmospheric seeing conditions.

\begin{figure}[bhtp]
\epsscale{0.75}
\plotone{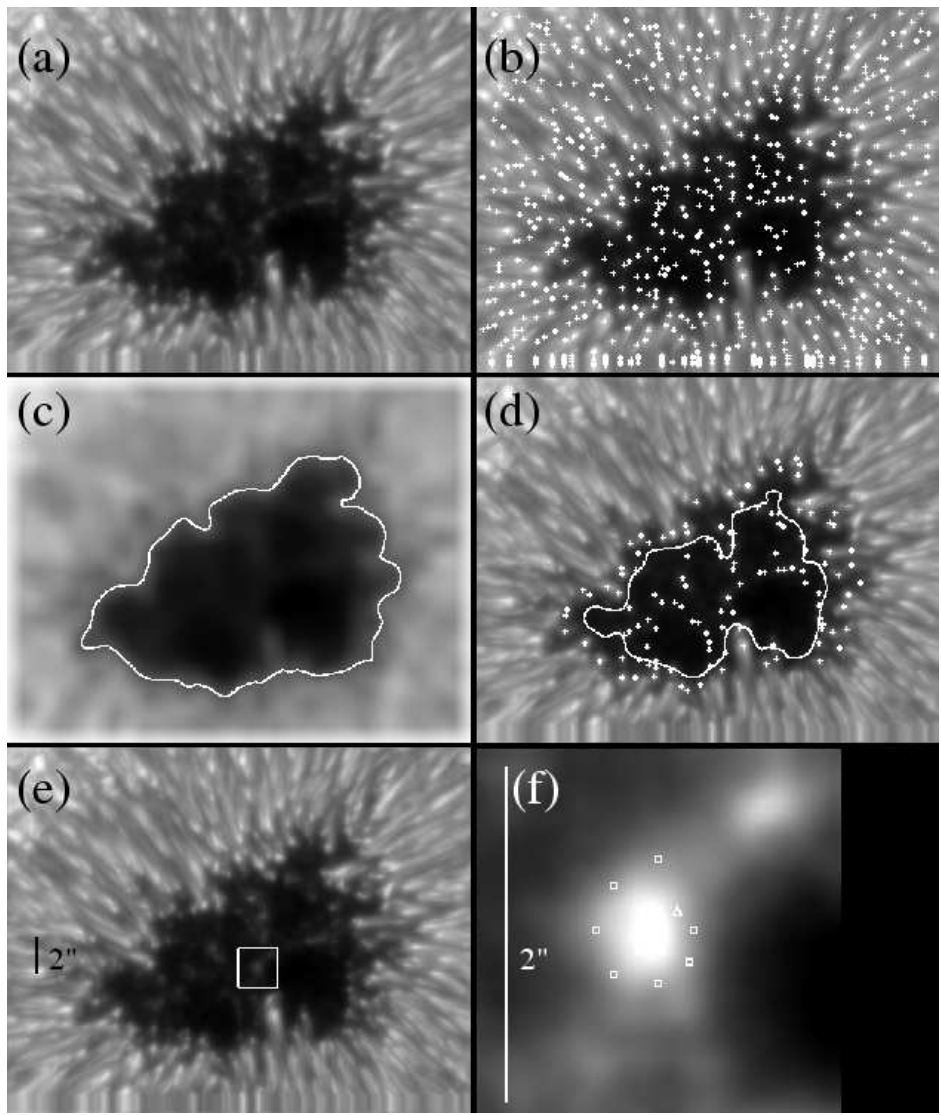}
   \caption{(a) Hanning filtered image (b) Peak detection (c) Background image. The contour denotes the boundary of the umbra ($I_{bg}$=0.4). (d) UD positions. The contour denotes the boundary of peripheral UDs ($I_{bg}=$0.2). UDs outside of the contour are peripheral, and UDs inside are central. (e) The position of the UD shown in (f). (f) An example of the size calculation. The 7 squares (threshold of $I(x,y)$=0.5*($I_{peak}(x,y)-I_{bg}(x,y)$)+$I_{bg}(x,y)$) and the triangle (inflection point) indicate the UD's boundary points for the 8 directions.}
    \label{fig:detection_algorithm}
\end{figure}

The algorithm consists of 6 steps. 
\begin{enumerate}
\item Identify local peaks where $I(x,y)$ ($I$: intensity of a pixel normalized by the averaged quiet region) is equal to the maximum value within $\pm$2 pixels ($\pm$78 km) of the vicinity. (Fig. \ref{fig:detection_algorithm}(b))
\item Construct background image ({\it bg}).  First, from the original image, we assign the minimum intensity within $\pm$6 pixels ($\pm$235 km) of its vicinity into each pixel.  The obtained image is boxcar-smoothed with a width of 20 $\times$ 20 pixels (782 km).  Define the umbra as the region where the intensity of {\it bg} is less than 0.4. This umbra region will be analyzed in the following steps. (Fig. \ref{fig:detection_algorithm}(c))
\item Define UD's positions as pixels where $I_{peak}(x,y)/I_{bg}(x,y)$ is larger than 1.3. The total number of UDs in each frame is on average 124. (Fig. \ref{fig:detection_algorithm}(d))
\item Calculate the size of the UDs. When determining the size, we apply two different methods and take the smaller value of the two. The first method uses the distance from the UD's peak position to the nearest inflection point; the second one uses the distance to the threshold of 0.5*($I_{peak}(x,y)-I_{bg}(x,y)$)+$I_{bg}(x,y)$. These calculations are applied to the 8 directions spaced at 45$^{\circ}$ intervals from the peak positions. Finally, we take the median of the 8 values. (Fig. \ref{fig:detection_algorithm}(e), (f))
\item Once the position of each UD in every frame has been decided, the temporal succession of each UD is determined as follows: the succession is confirmed if an UD is found in the next frame (taken 25\,s later) within $\pm$2 pixels of its previous position. If no UD is found, the continuation ends at that point. However, when there is no UD in the next frame but an UD is found within $\pm$2 pixels in the following one or two frames, the succession is continued. Fission (two UDs within $\pm$2 pixels) and fusion (two UDs coalesce into one UD) events are taken into account, though these events are scarce (less than 0.2\% of all UDs in one frame). This procedure is applied to all UDs until they fade out, or the final frame of the data is reached.
\item Classify UDs into two categories, i.e., central UDs (umbral origin) and peripheral UDs (penumbral origin). The boundary between central and peripheral UDs is set to the contour line of $I_{bg}$=0.2. Peripheral UDs are those which have their origins in brighter region, i.e., outside of the $I_{bg}$=0.2 contour. (Fig. \ref{fig:detection_algorithm}(d))
\end{enumerate}

Using the above method, 2268 UDs were detected.
Out of 2268 UDs, 825 of them were central UDs and 1443 of them were peripheral UDs.
We want to note that 245 UDs either already existed before the first frame or lasted longer than the end of the observation.
These 245 UDs can not be tracked from the beginning to the end, so their lifetimes are underestimated.
Nevertheless, we include these UDs in our analysis, because they sometimes represent long-life UDs.

\section{Results}
We detected the temporal trajectories of 2268 UDs using the method described in {\S3}.
Using this information, we calculated the 8 parameters which characterize the UDs: lifetime, average size, brightness ratio, velocity amplitude, velocity orientation, magnetic field strength, field inclination, and field azimuth. The term ``velocity'' in our analysis refers to the proper motion, that is, the horizontal velocity. These parameters are defined as follows:

\begin{itemize}
\item lifetime ($T$) $\cdots$ (temporal cadence 25\,s) $\times$ (number of frames in which the UD is observed)
\item average size ($S$) $\cdots$ the average of the individual size calculated at frames in which the UD is observed (the size in each frame is described in {\S}3, step [4])
\item brightness ratio ($R$) $\cdots$ the average of the individual $I_{peak}/I_{bg}$ calculated at frames in which the UD is observed
\item velocity amplitude ($|V|$) $\cdots$ (the distance between its emergence and fade-out location)/(lifetime)
\item velocity orientation ($v$) $\cdots$ the direction from its emergence location to its fade-out location measured counter-clockwise from the left-to-right (west), i.e., in the opposite direction of the field azimuth.
\item magnetic field strength ($B$) $\cdots$ the magnetic field strength at its origin (note that the magnetic field information is global, composed from two SP maps taken at different times. See {\S}2)
\item field inclination ($i$) $\cdots$ the field inclination at the UD's origin
\item field azimuth ($\psi$) $\cdots$ the field azimuth at the UD's origin
\end{itemize}   

\subsection{Histogram}
\begin{figure}[bhtp]
\epsscale{0.9}
\plotone{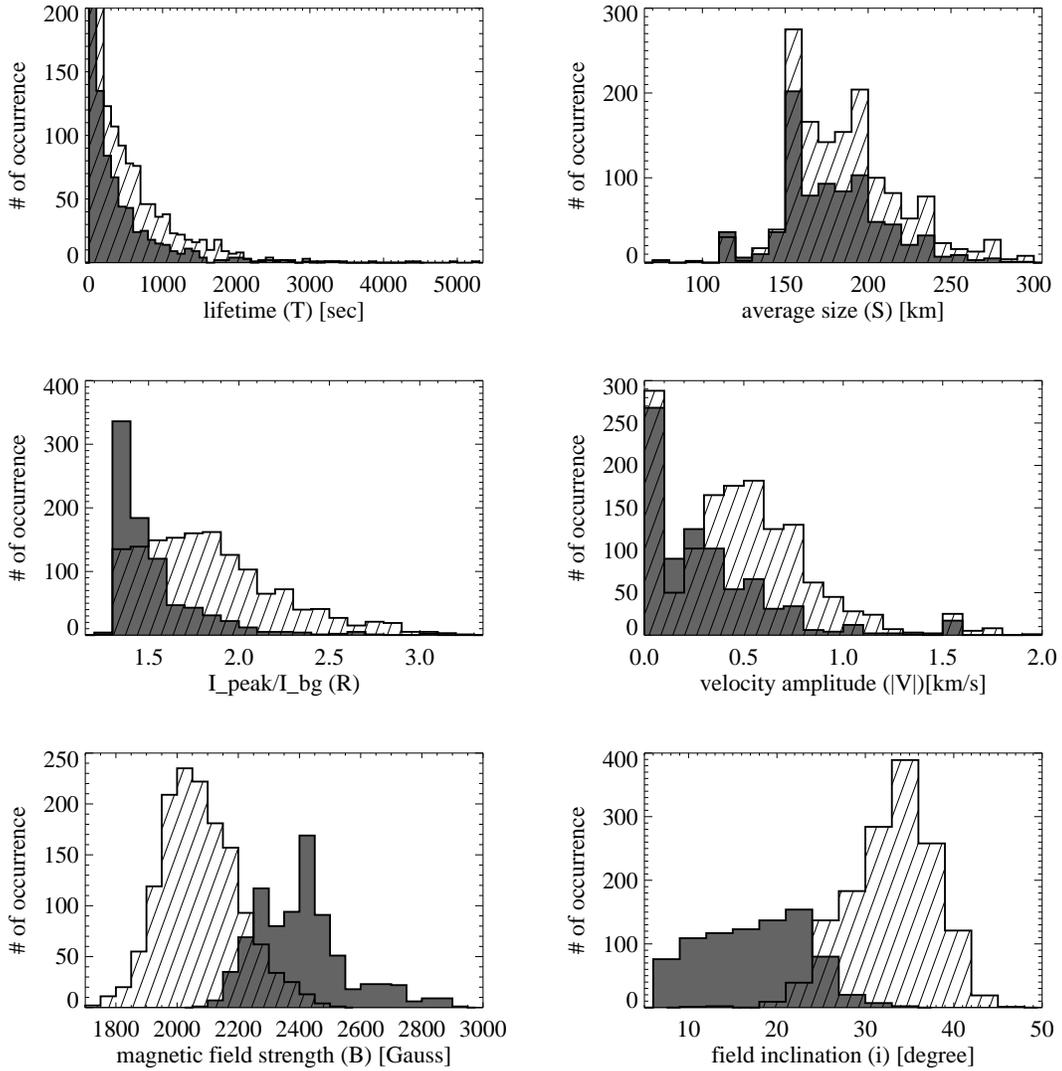}
   \caption{Histograms for 2268 UDs. The gray and hatched regions correspond to the central and peripheral UDs, respectively. 
   }
    \label{fig:histogram_parameter}
\end{figure}

\begin{table}
\begin{center}
\caption{Average of UD parameters \label{tbl:average_parameter}}
\begin{tabular}{lccc} \\
\tableline
 & Average of  & Average of  & Average of \\
 & 2268 UDs & 825 central UDs & 1443 peripheral UDs \\
\tableline
$T$ [sec] & 441 & 391 & 469 \\
$S$ [km] & 184 & 178 & 187 \\
$R$ & 1.73 & 1.51 & 1.85 \\
$|V|$ [km s$^{-1}$] & 0.44 & 0.33 & 0.50 \\
$B$ [Gauss] & 2197 & 2406 & 2077 \\
$i$ [degree] & 27 & 17 & 33 \\
\tableline
\end{tabular}
\end{center}
\end{table}

Figure \ref{fig:histogram_parameter} shows the histograms of the 6 parameters ($T, S, R, |V|, B, i$).
The gray and hatched bars indicate the central and the peripheral UDs, respectively.
The average values of each parameter are summarized in Table\,\ref{tbl:average_parameter}.
As the diffraction limit of blue continuum imaging is 164\,km, UDs with a size smaller than 164\,km can not be defined accurately.
We found that peripheral UDs tend to have brighter intensity, faster proper motion, weaker field with a larger field inclination than central UDs.
This result is consistent with \cite{Sobotka1997a,Sobotka1997b} and \cite{Kitai2007}, which supports the validity of our automatic detection method.

\subsection{Scatter Diagrams}
Next, we study the scatter plots of the parameters of UDs against the magnetic field strength and inclination, as shown in Fig.\,\ref{fig:scatter_parameter}.
Please refer to Fig.\,\ref{fig:SPbefore_after} for the spatial distributions of the magnetic field components.
Considering the large 1$\sigma$ errors, the lifetimes of UDs are independent of the magnetic field strength and inclination.
The sizes of UDs are almost constant, at approximately 190 km. 
There is a hint that the average size in the dark umbral core ($B>2600$\,Gauss and $i\sim20^{\circ}$) is abot 20\% smaller compared to 190\,km, though the statistical significance remains unproved.
Because of the large scatter and non-Gaussian distributions of the samples, the brightness ratio $I_{peak}/I_{bg}$ does not show a clear correlation with the field strength, although high contrast UDs seems to correspond to low magnetic field ($\sim$2000\,Gauss).  
We will study this topics further in \S{4.4}.
The velocity amplitude shows a clear dependence both on the field strength and on the field inclination, with the dependence on the field inclination being more pronounced than on the field strength. 
Most of the samples with $|V|$=0\,km s$^{-1}$ and $>$1.5 km s$^{-1}$ corresponds to short-lived UDs, because one pixel movement (39\,km) in one frame (25\,s) results in $|V|$=1.56\,km s$^{-1}$.
If we neglect these short-lived UDs, the velocity amplitude clearly increases with the field inclination.

Figure\,\ref{fig:velori_scatter} shows the 6 scatter diagrams of the velocity orientation of UDs versus the field azimuth for 1762 UDs for different ranges of the field inclination.
Here we ignored the 506 UDs which had no detectable movement from their emergence to fade-out, because their velocity orientation could not be defined.
Note that the 0$^{\circ}$ direction of the velocity orientation and of the field azimuth have opposite directions.
Thus, an UD which move toward the center of the umbra along the field line has the same velocity orientation angle as the field azimuth.
In the bottom panels of Fig.\,\ref{fig:velori_scatter}, the velocity orientation angles are nearly parallel to the field azimuth.
This is because the peripheral UDs generally move umbra inward.
However, this correlation is weaker for UDs with smaller field inclinations (upper panels in Fig.\,\ref{fig:velori_scatter}).
The correlation coefficients for the different inclination ranges are: 0.22 (inclination$<$15$^{\circ}$), 0.45 (15$^{\circ}$-20$^{\circ}$), 0.47 (20$^{\circ}$-25$^{\circ}$), 0.63 (25$^{\circ}$-30$^{\circ}$), 0.64 (30$^{\circ}$-35$^{\circ}$), and 0.75 ($>$35$^{\circ}$).
Therefore, we conclude that the velocity orientation of UDs in regions with large field inclination is determined by the field azimuth, while the velocity orientation of UDs in regions with small field inclination is weakly dependent on the field azimuth.

\begin{figure}[bhtp]
\epsscale{0.85}
\plotone{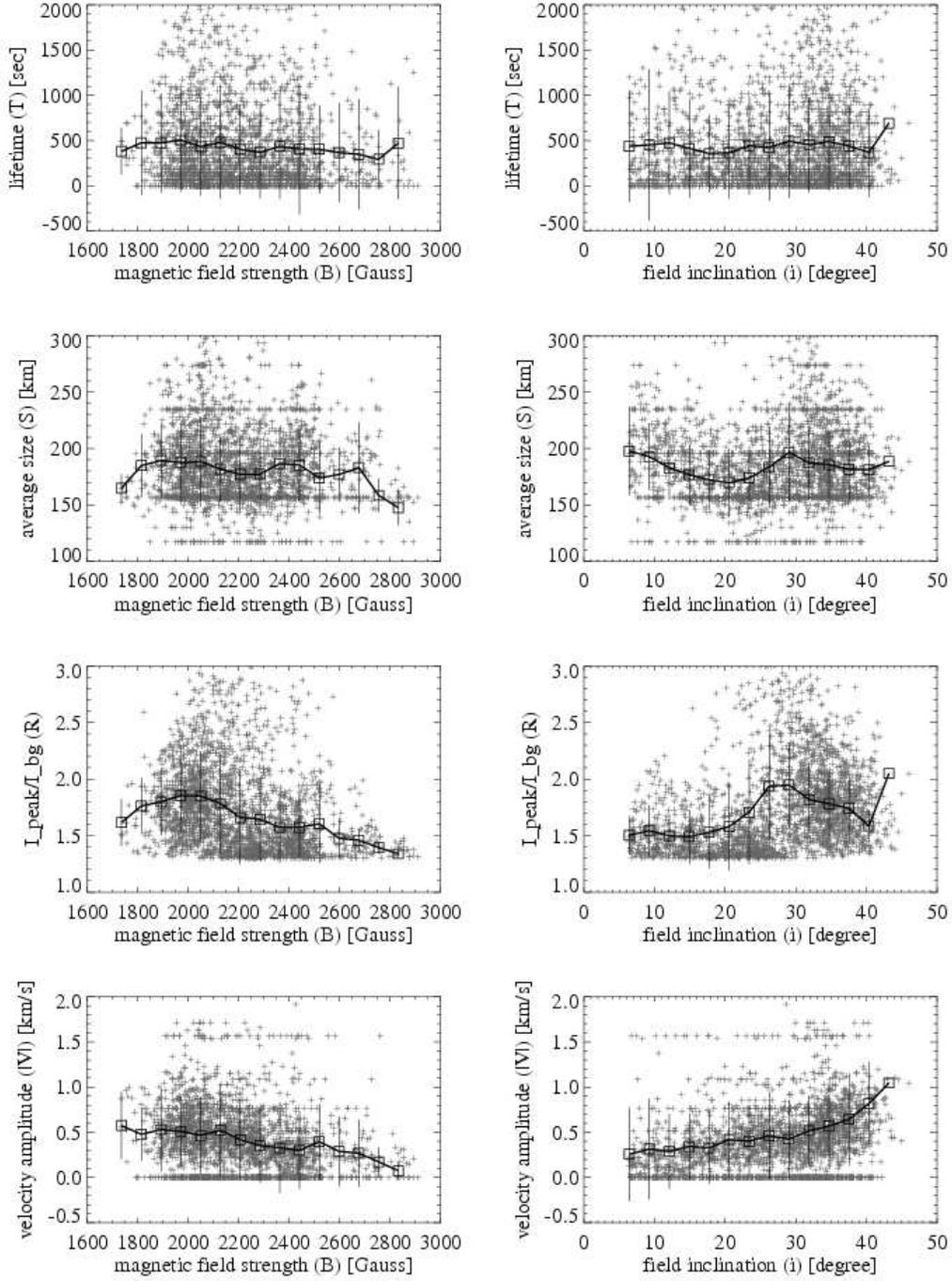}
   \caption{{\it Left column:} Scatter plots of the lifetime, average size, $I_{peak}/I_{bg}$, and velocity amplitude versus the magnetic field strength. The average of bins of 80\,Gauss is shown with square symbols, and solid lines denote the standard deviation error bars. {\it Right column:} Scatter plots of the lifetime, average size, $I_{peak}/I_{bg}$, and velocity amplitude versus the field inclination. The average of bins of 3$^{\circ}$ is shown with square symbols, and solid lines denote the standard deviation error bars.}
    \label{fig:scatter_parameter}
\end{figure}
	
\begin{figure}[bhtp]
\epsscale{1.0}
\plotone{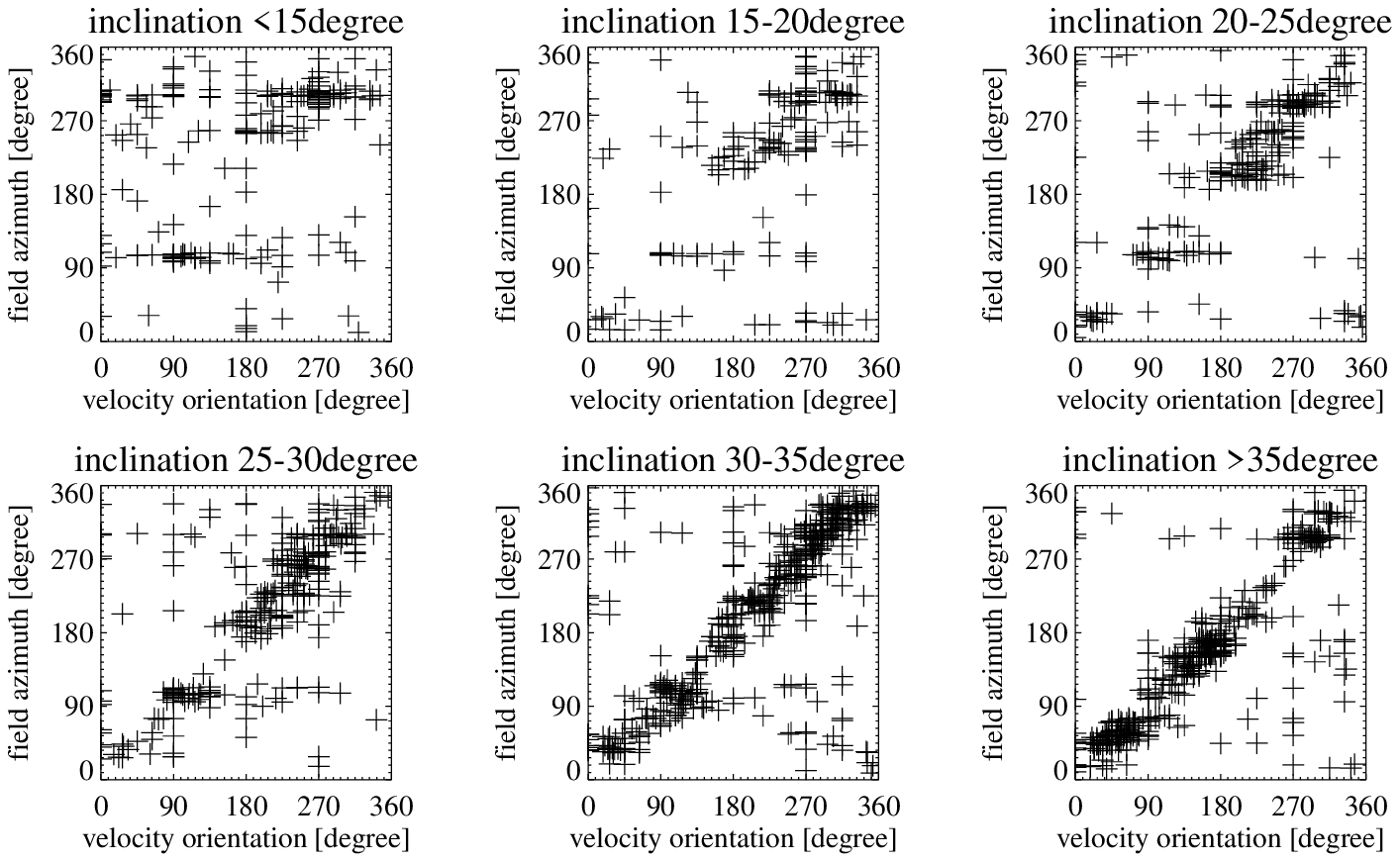}
   \caption{Scatter plots of the velocity orientation versus the field azimuth for different ranges of the field inclination}
    \label{fig:velori_scatter}
\end{figure}

\subsection{Spatial Distribution}
It is known that dark cores include few UDs \citep{Beckers1968, Kitai2007}, because the magnetic field in dark cores is too strong to produce effective magnetoconvection.
We confirmed that strongly magnetized regions include fewer UDs, as shown in Fig.\,\ref{fig:magnetic_density}.
The occurrence rate is given by the ratio of the number of UDs within regions of a specific field strength to the area of these regions.
The left image in Fig.\,\ref{fig:magnetic_density} shows the distribution of the emergence positions of all UDs.
A cluster of UDs appear in the areas surrounding the dark cores, which may be an indication of flux separation \citep{Tao1998, Weiss2002}.
As well as UDs surrounding the dark cores, we can see some cellular patterns in the left image of Fig.\,\ref{fig:magnetic_density}.
These cellular patterns may reflect the global subsurface magnetic field configuration as was supposed in the cluster model \citep{Parker1979}.
The temporal evolution of these global cellular patterns is of great interest and should be studied further.

\begin{figure}[bhtp]
\epsscale{1.0}
\plotone{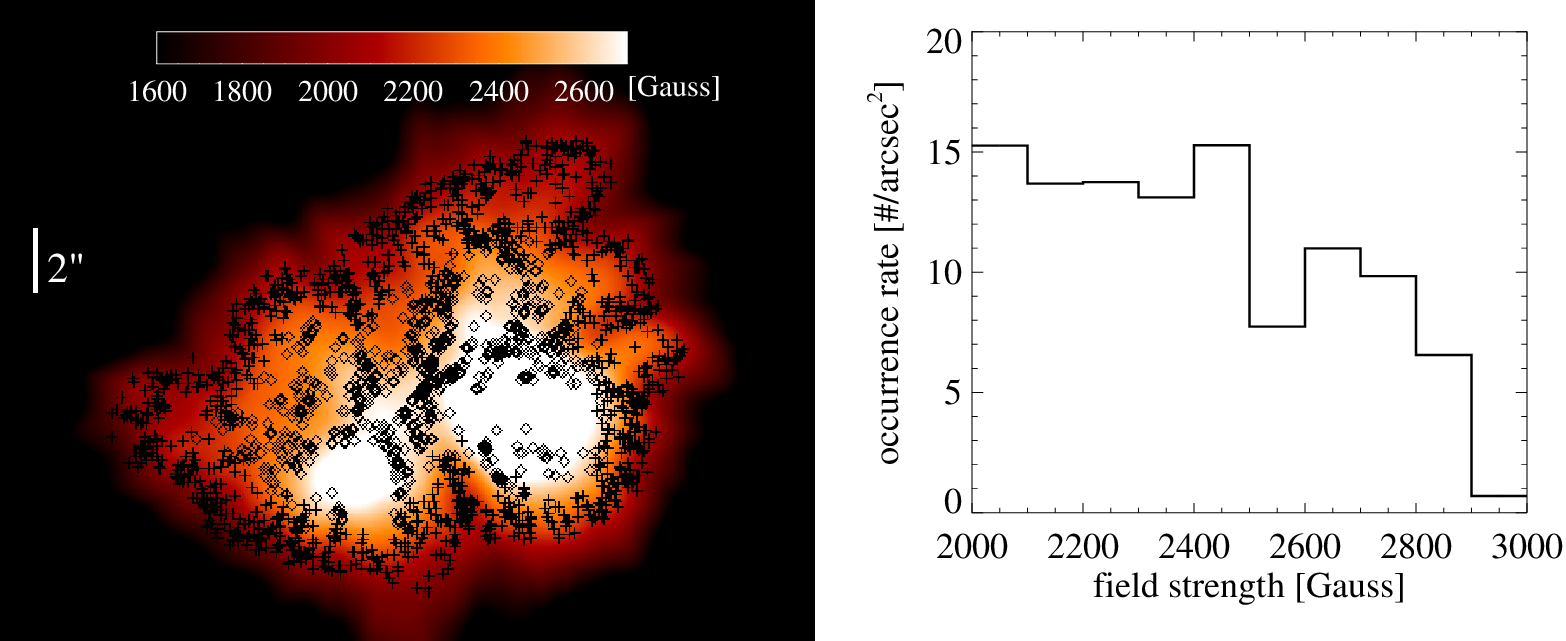}
   \caption{{\it Left}: The emergence positions of the central UDs (diamonds) and the peripheral UDs (plus signs), overlaid on the background image of the magnetic field strength. {\it Right}: The UD's occurrence rate versus the field strength with field strength bins of 100 Gauss.}
    \label{fig:magnetic_density}
\end{figure}

\begin{figure}[bhtp]
\epsscale{1.0}
\plotone{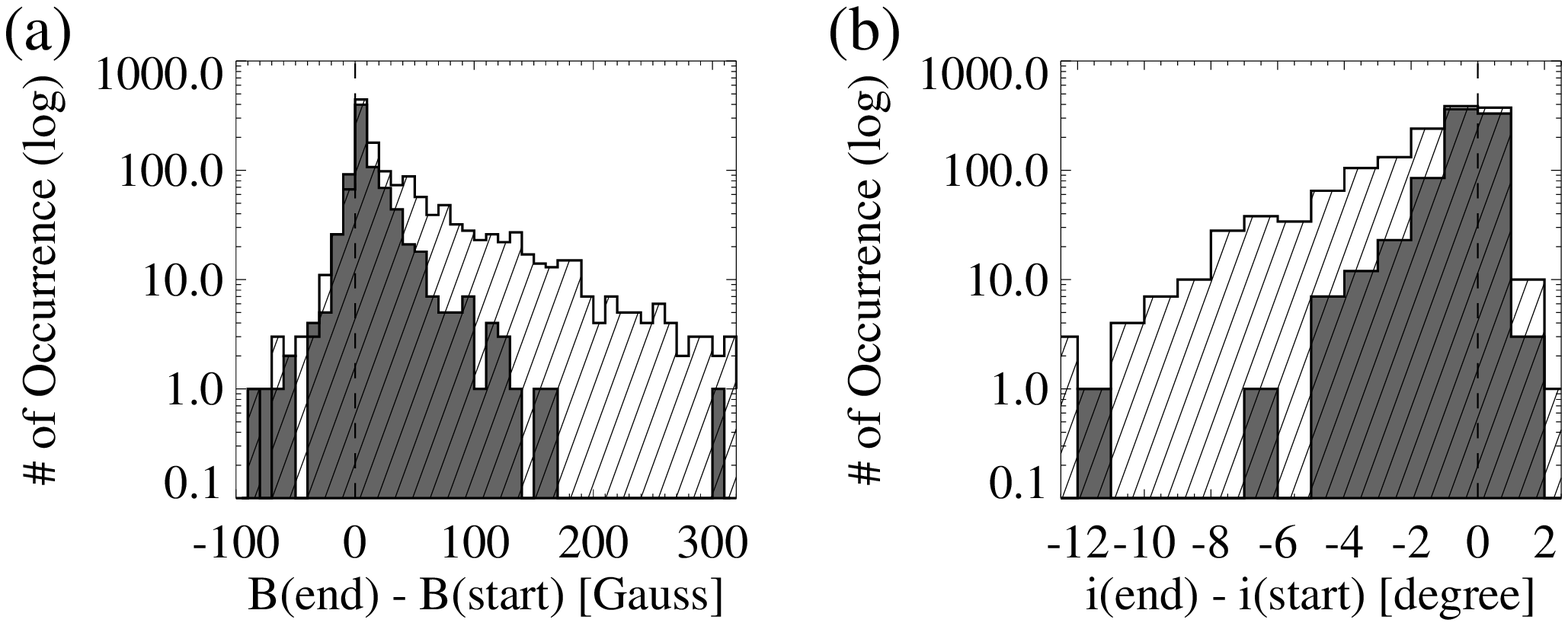}
   \caption{(a) Histogram of the field strength differences between their emergence (start) and fade-out (end) locations (log scale). The gray bars indicate central UDs, the hatched bars indicate peripheral ones. (b) Histogram of the field inclination differences between their emergence (start) and fade-out (end) locations (log scale). The gray bars indicate the central UDs, and the hatched bars indicate the peripheral ones.}
    \label{fig:magnetic_UD}
\end{figure}

The histograms of the difference between the emergence (start) and fade-out (end) locations of the field strength and the inclination are shown in Fig.\,\ref{fig:magnetic_UD}.
73\% of the peripheral (hatched) and 56\% of the central UDs (gray) faded out in an area with a larger field strength than in their emergence location, while only 8.0\% of the peripheral and 16\% of the central UDs faded out in an area with a smaller field strength.
Similar results are found for the field inclination in Fig.\,\ref{fig:magnetic_UD}(b):
73\% of the peripheral and 58\% of the central UDs faded out in an area with a less inclined field than in their emergence location, while 7.5\% of the peripheral and 12\% of the central UDs faded out in an area with a more inclined field.
This means that not only the peripheral UDs, but also the central UDs are likely to appear in weaker field regions with more inclined field orientation and disappear in stronger field regions with less inclined field orientation.
This is a natural consequence for peripheral UDs, because they move inward from the penumbra to the umbra. 
For central UDs, however, the mechanism is less obvious.
Further investigations should be possible using the temporal evolution of the magnetic field.

\subsection{Lightcurve}
\begin{figure}[bhtp]
\epsscale{0.8}
\plotone{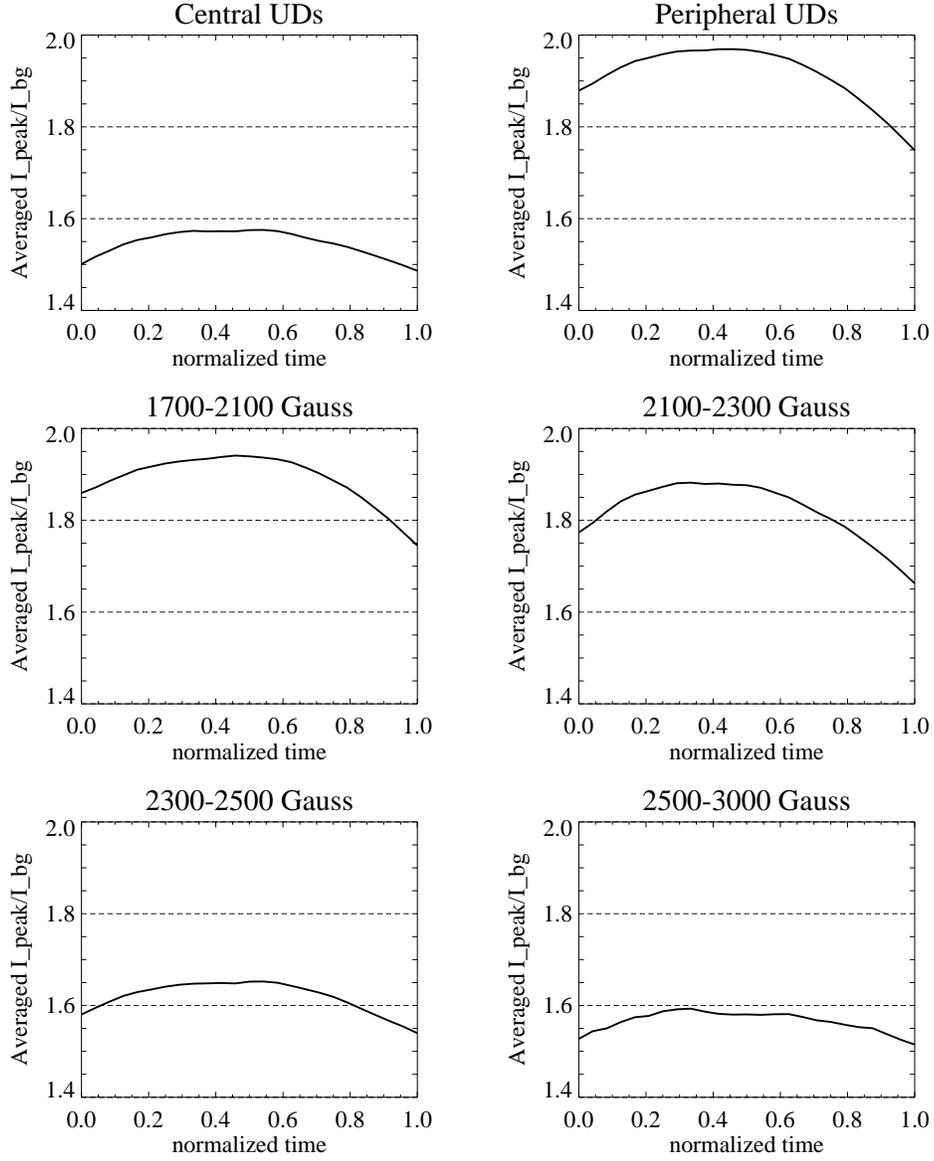}
   \caption{Averaged lightcurves of the central UDs ({\it top left}), the peripheral UDs ({\it top right}), UDs born in 1700-2100 Gauss regions ({\it middle left}), UDs born in 2100-2300 Gauss regions ({\it middle right}), UDs born in 2300-2500 Gauss regions ({\it bottom left}), and UDs born in 2500-3000 Gauss regions ({\it bottom right}).}
    \label{fig:lightcurve_total}
\end{figure}

We compared the characteristic lightcurves of UDs in different field strength regions. 
Only UDs with lifetimes of $T$$>$120 s are selected for analysis in this section.
First, UD lifetimes are normalized to unity (0 at their emergence and 1 at their fade-out).
A typical lightcurve is then defined as the average of the temporal variation of the UD brightness.
The results are shown in Fig.\,\ref{fig:lightcurve_total}.
The characteristic lightcurves of the central and the peripheral UDs (top panels of Fig.\,\ref{fig:lightcurve_total}) are consistent with those in \cite{Kitai2007} and \cite{Riethmuller2008}.
The uniqueness of our analysis lies in the lower 4 plots in Fig.\,\ref{fig:lightcurve_total}.
We notice a clear dependence of the contrast and the amplitude of the brightess fluctuations on the magnetic field strength.  
High contrast UDs appear in low magnetic field regions, as was suggested in \S{4.2}.
In the strong field bands, the amplitudes of fluctuations of the brightness ratio get smaller, and the lightcurves show symmetric brightening and darkening.
In the weak field bands, the amplitude of the brightness fluctuations is large, and the lightcurves show fast brightening and slow darkening.

\begin{figure}[bhtp]
\epsscale{0.8}
\plotone{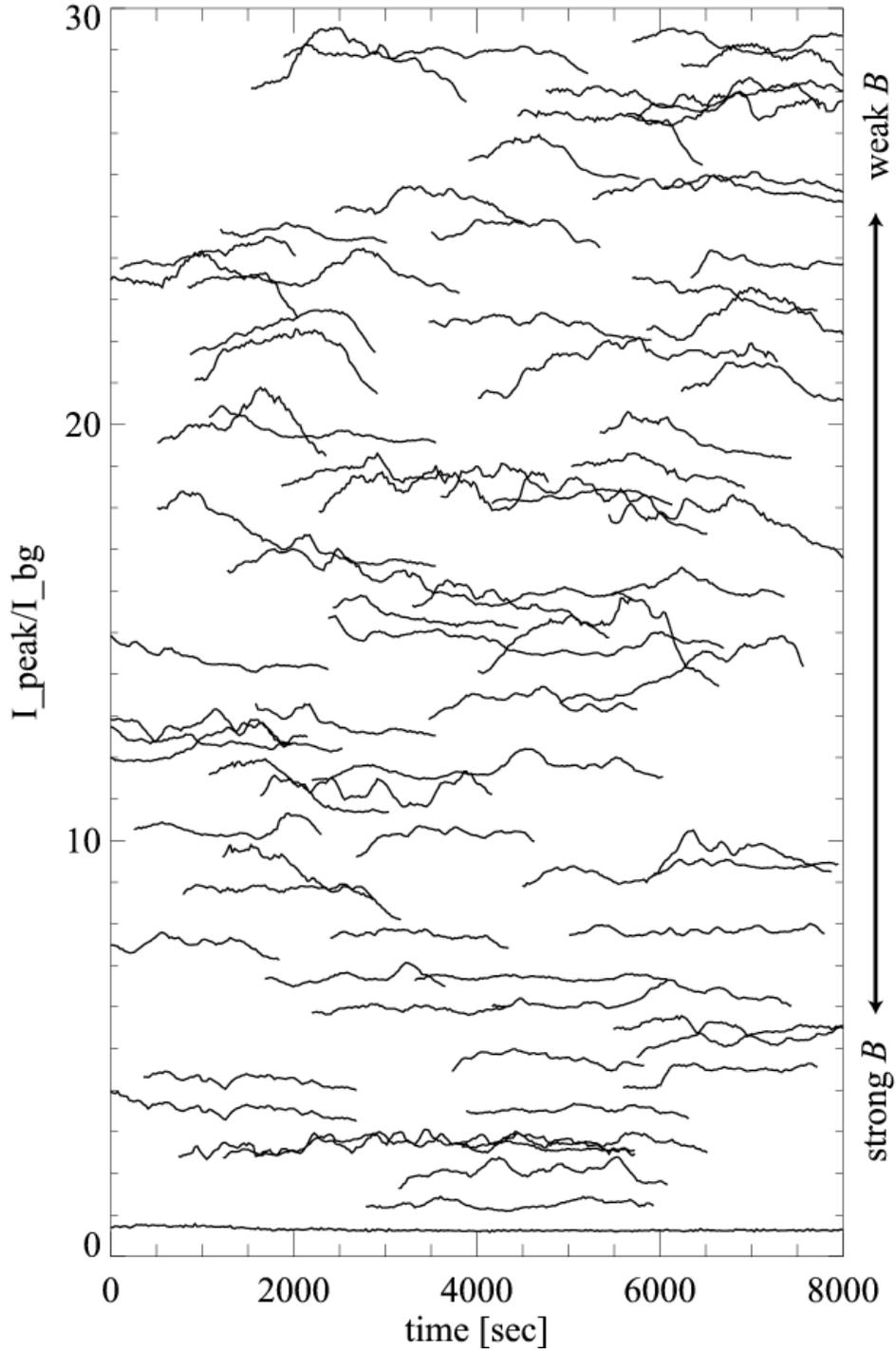}
   \caption{Lightcurves of 76 UDs with $T$$>$1800 s. The lightcurves are arranged by their field strengths from top to bottom, i.e., the upper curves have the weaker field strength. The lightcurves are spaced by constant vertical offsets. The bottommost lightcurve is that of a point in the dark core.}
    \label{fig:longlife_lightcurve}
\end{figure}

Another important property of the UD's lightcurve is its oscillation. 
We show the lightcurves of 76 UDs with $T$$>$1800 s in Fig.\,\ref{fig:longlife_lightcurve}.
The lightcurves are spaced from each other with a constant offset.
We arranged the lightcurves in order of their field strengths from top to bottom.
The bottommost lightcurve is the brightness of a fixed point in the dark core, which may represent the error fluctuation level.
The lightcurves display oscillatory fluctuations.
We should keep in mind that it is very difficult to confirm whether this lightcurve is the true oscillation in one UD, or the conglomerates of multiple UDs occurring side by side.
The identification threshold described in \S{3} may also affect the result.

To find the characteristic frequency of the oscillations, a Fourier transformation analysis was performed.
The procedure for the Fourier transformation analysis is as follows:
firstly, we applied the third order polynomial fit to each lightcurve, and subtracted it from the original lightcurve to remove the gradual variations.
Secondly, the IDL routine {\it FFT} (Fast Fourier Transformation) is applied to the subtracted lightcurves.
The amplitude spectrum of the lightcurve is calculated by taking the absolute value of the Fourier transformed function.
Figure \ref{fig:longlife_power} shows the amplitude spectra of 76 lightcurves in color scale.
The UDs are arranged in order of their field strength, i.e., UD 0 corresponds to the strongest field (2740 Gauss) and UD 75 corresponds to the weakest field (1880 Gauss).
As is shown in Fig.\,\ref{fig:lightcurve_total}, the amplitude of fluctuation of the UD brightness is greater in regions with weaker fields than those with stronger fields.
The lower frequency components at 1-2mHz (8-16 min) are dominant for almost all of the UDs.
Note that, since the shortest lifetime of the analyzed UDs is 1800 s, the detectable frequency is limited to larger values than 1.1 mHz.
The time scale of 8-16 min is consistent with the typical lifetime of UDs.
This characteristic frequency was not found in the lightcurve of the dark core.  
There is no systematic variation in the amplitude spectra from the weaker magnetized area to the stronger magnetized area.
The constant lifetime shown in Fig.\,\ref{fig:scatter_parameter} is consistent with this result.

 \begin{figure}[bhtp]
\epsscale{0.8}
\plotone{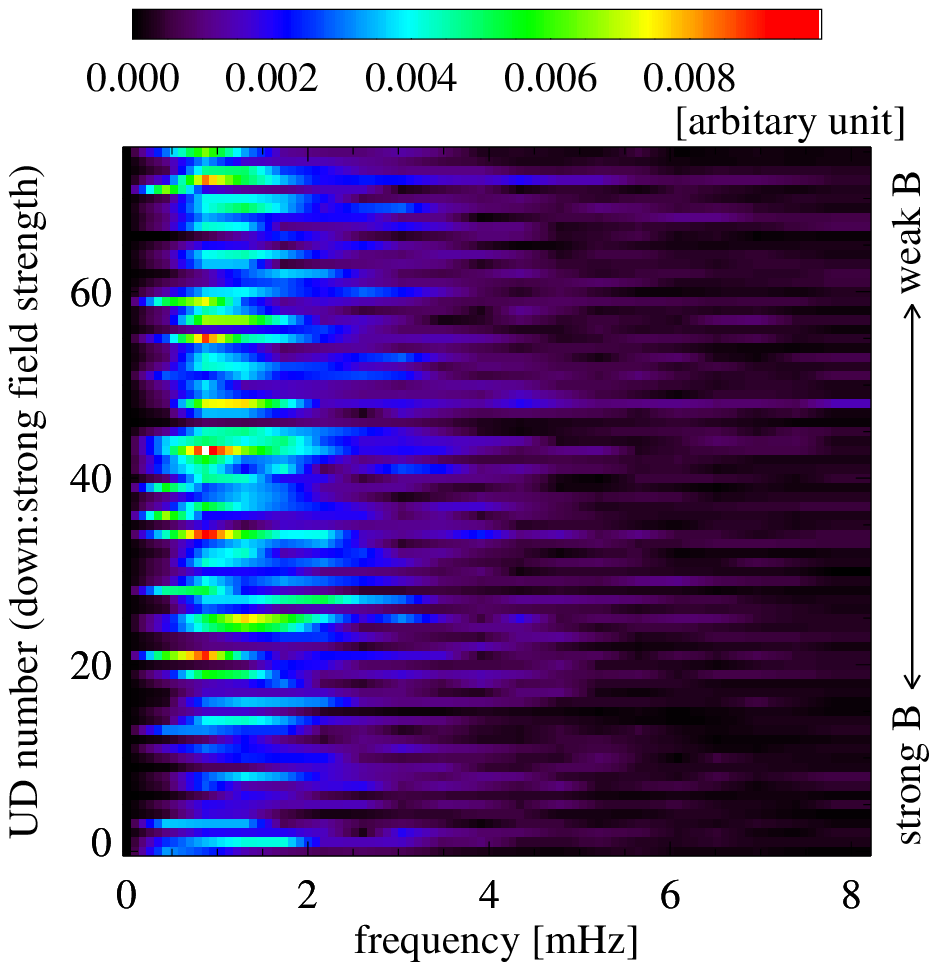}
\caption{Fourier amplitude spectra of the lightcurves of 76 UDs. The horizontal axis indicates the frequency, the vertical axis the arrangement of the 76 UDs (lower positions correspond to stronger fields).}
    \label{fig:longlife_power}
\end{figure}

\section{Discussion}	
We analyzed in detail the relationships between the UD's parameters and the magnetic structure.
The precise measurement of the magnetic field and the stable high-resolution imaging by the \emph{Hinode} SOT made our analysis possible.
 
The obtained distributions of the lifetime, the average size, and the proper motion of UDs confirm the performance of our automatic detection algorithm.
The averages of each parameter for the central and the peripheral UDs are listed in Table\,\ref{tbl:average_parameter}.
Using these parameters (the lifetime, average size, brightness ratio, velocity amplitude, and velocity orientation) plus three magnetic field components (the field strength, field inclination, and field azimuth), we are able to obtain important correlations between UDs and the magnetic field.
The results can be summarized as follows:

\begin{enumerate}
\item The lifetimes and average size of UDs show almost no dependence on both the magnetic field strength and field inclination.
\item Highly contrasted UDs come up in the zone near the penumbra, where the magnetic field  is weak ($\sim$2000\,Gauss) and inclined above 30 degree.
\item The velocity amplitudes of the UDs are correlated with the field inclination.
\item The velocity orientations of the UDs in the larger field inclination regions are nearly parallel with field azimuth, while those of the UDs in the smaller field inclination areas have virtually no or random proper motion.
\item The UDs tend to have their origins in regions with weaker, more inclined magnetic fields, and to disappear in regions with stronger, less inclined magnetic fields.
\item There is a negative correlation between the field strength and the occurrence rate of UDs.
\item The analysis of the oscillations of the UD's lightcurves reveals that low frequency components at 1-2 mHz (8-16 min) are dominant for all of the UDs, regardless of the magnetic field strength.
\end{enumerate}

The scatter diagrams shown in Fig.\,\ref{fig:scatter_parameter} have large 1$\sigma$ errors and the conclusion is not straightforward in our study. 
The lifetime and average size of UDs are almost independent on the field strength, though there is a hint that the average size at the dark cores is about 20\% smaller compared to that at 2000 Gauss.  
The most significant feature is the positive correlation between the field inclination and UD's velocity amplitude.
The gappy model \citep{Spruit2006, Heinemann2007} can explain the proper motion of the peripheral UDs.
When hot gas ascends along the inclined magnetic field, the gas undergoes radiative cooling and becomes denser. 
The heavy gas bends the surrounding magnetic field line, increasing their inclination, and then the magnetic field strength at the upper side of the bent field lines is weakened. 
In order to balance the reduced magnetic pressure, more hot gas rises up from below and causes the apparent movement of the UD. 
This process occurs repeatedly, and produces the inward migration of peripheral UDs \citep{Scharmer2008}.
On the other hand, this ``bending" process will not occur if the field line is vertical. This may be the reason why the central UDs do not show systematic inward migration. 
The larger correlation coefficient between the field azimuth and the UD's velocity orientation in the region with larger field inclination provides further evidence of this behavior.
However, it is difficult for this idea to explain the fact that the lifetime of central and peripheral UDs are the same.
Another possible mechanism is the moving tube model \citep{Schlichenmaier2002}, in which the inward migration corresponds to the footpoint of a rising flux tube.
Both models predict the inward migration for UDs in more inclined fields.
  
\cite{Weiss2002} explained UDs as small scale magnetoconvection modified by the existence of strong magnetic fields.
They also suggested that vigorous convection should take place in the weakly magnetized area, which may correspond to light bridges.
The small scale convection is a natural consequence of the theoretical model for suppressed eddy motion by strong magnetic fields \citep{Weiss1981, Blanchflower2002}.
In our case, the average size of UDs are almost constant $\sim$190\,km, though there is a hint that the average size in the dark umbral cores ($B>2600$\,Gauss and $i\sim20^{\circ}$) is a little smaller.
The statistical analysis of the size of UDs is strongly affected by the definition of their size.
A more accurate comparison is only possible by a common analysis of both the observed and the simulated umbra smeared using a point spread function consistent with observations.
The constant lifetimes we found for UDs is inconsistent with the lifetimes of the simulated UDs by Sch\"ussler \& V\"ogler (private communication).
The mean lifetime of the simulated UDs is 34\,min for 2000\,Gauss sunspots, 28\,min for 2500\,Gauss sunspots, and 25\,min for 3000\,Gauss sunspots, which is significantly longer than the average lifetime of our analysis (7.3 min). 

The oscillation analysis of the UD's lightcurves was first carried out by \cite{Sobotka1997b}.
However, they used ground-based, seeing-affected observation data.
The \emph{Hinode} data is free from such variable atmospheric conditions, which greatly extends the reliability of the oscillation analysis (see Fig. 8 in Watanabe et al. 2009).
The strong signals at 1-2mHz (8-16 min) are commonly found for the UDs with $T>$ 40 min.
As the typical lifetime of UDs is of the order of 10 mim, this result may show that the successive appearance of UDs causes the oscillatory lightcurve. 
We could not find any systematic change in the amplitude spectra of the UDs in the weaker magnetized area and those in the stronger magnetized area.  
This is consistent with the common lifetime in weak and strong magnetic fields.
However, the interpretation of the lightcurve oscillations of an UD is not straightforward, because they can be caused either by overlapping with adjacent UDs, by an additional heat flux into the UD, or by the period of the oscillatory magnetoconvection.

Recent 3D MHD simulations have successfully simulated the basic characters of UDs.
The observational properties of UDs found in this study provide a good test for the theory of UDs, and they should be investigated in future numerical simulations. 
This strategy will lead to a more realistic understanding of the structure of sunspots.

\acknowledgments
We thank Manfred Sch\"ussler (Max Planck Institute for Solar System Research) and Hiroaki Isobe (Unit of Synergetic Studies for Space) for helping us with the interpretation of the lifetime and distribution of UDs.  We are grateful to all the staff of Hida Observatory, and to all staff and students of Kyoto University and Kwasan and Hida Observatories for fruitful discussions.  The authors are partially supported by the grant-in-aid for the Global COE program ``The Next Generation of Physics, Spun from Universality and Emergence" from the Ministry of Education, Culture, Sports, Science and Technology (MEXT) of Japan, by the grant-in-aid for `Creative Scientific Research The Basic Study of Space Weather Prediction' (17GS0208, PI: K. Shibata) from the Ministry of Education, Science, Sports, Technology, and Culture of Japan, and by the grant-in-aid for the Japanese Ministry of Education, Culture, Sports, Science and Technology ( No.19540474).  Hinode is a Japanese mission developed and launched by ISAS/JAXA, with NAOJ as domestic partner and NASA and STFC (UK) as international partners.  It is operated by these agencies in co-operation with ESA and NSC (Norway).

\end{document}